\documentclass[10pt]{article}
\usepackage{multicol}
\usepackage{graphicx}
\usepackage{amsmath}

\begin{document}

\begin{center}
{\Large \bf Decoupling of non-strange, strange and multi-strange particles from the system in Cu-Cu, Au-Au and Pb-Pb collisions at high energies}

\vskip1.0cm

M.~Waqas$^{1,}${\footnote{Corresponding author. Email (M.Waqas):
waqas\_phy313@yahoo.com; waqas\_phy313@ucas.ac.cn}},G. X. Peng$^{1,2,3}$ {\footnote{Corresponding author. Email (G. X. Peng): gxpeng@ucas.ac.cn}},
Z. Wazir $^{4}$ {\footnote{author. Email (Z.wazir): zafar\_wazir@yahoo.com}}

\vskip.25cm

{\small\it $^1$ School of Nuclear Science and Technology, University of Chinese Academy of Sciences,
Beijing 100049, China,

$^2$ Theoretical Physics Center for Science Facilities, Institute of High Energy Physics, Beijing 100049, China,

$^3$ Synergetic Innovation Center for Quantum Effects \& Application, Hunan Normal University, Changsha 410081, China

$^4$ Department of Physics, Ghazi University, Dera Ghazi khan, Pakistan}
\end{center}

\vskip1.0cm

{\bf Abstract:} Transverse momentum spectra of the non-strange, strange and
multi-strange particles in central and peripheral Copper-Copper, Gold-Gold
and Lead-Lead collisions are analyzed by the blast wave model with Boltzmann
Gibbs statistics. The model results are approximately in agreement with
the experimental data measured by BRAHMS, STAR, SPS, NA 49 and WA 97 Collaborations
in special transverse momentum ranges. Bulk properties in terms of kinetic freeze out temperature, transverse
flow velocity and freezeout volume are extracted from the transverse momentum spectra of the particles.
Separate freeze out temperatures are observed for the non-strange, strange and multi-strange
particles which maybe due to different reaction cross-sections of the interacting
particles and it reveals the triple kinetic freezeout scenario  in collisions at
BRAHMS, STAR, SPS, NA 49 and WA 97 Collaborations, however the transverse flow velocity and freezeout volume
are mass dependent and they decrease with the increasing the rest mass of the particles.
Furthermore, the kinetic freezeout temperature, transverse flow velocity and kinetic freezeout volume in central
nucleus-nucleus collisions are larger than those in peripheral collisions. Besides, the larger
kinetic freezeout temperature and freezeout volume are observed in the most heaviest nuclei collisions, indicating their dependence on the size of interacting system.
\\



\vskip1.0cm
\begin{multicols}{2}

{\section{Introduction}}

Heavy ion collisions have been used to study
the Quark Gluon Plasma (QGP) phase transition into a hadronic gas [1--4]
for more than a decade. The phase transition was initially assumed to be
either first or second order [5--8], but the quantum chromodynamics
(QCD) calculations proved that it is a cross-over [9--11]. There is no
single critical temperature in a cross-over phase transition where a
phase transition occurs and all the degrees of freedom are going to be
switched between the phases. It is impossible to idiomatically define
(even though in the case of QCD possibly many exists) the characteristic
temperature ($T_c$) and it depends on what potential order parameter is
being observed. Chiral condensate, Polyakov loop and strangness suscepti-
bility [12]  were focused in the early lattice QCD calculations and found
a wide range in characteristic temperatures. The most recent lattice QCD
calculations has shown an approximate difference of 15-20 MeV in the hadronization
temperatures between the light and strange particles [13].

Temperature is one of the most important concept in high energy heavy-ion
collisions. Several temperatures namely the initial temperature, chemical
freezeout temperature, thermal or kinetic freezeout temperature and
the effective temperature are frequently used in the physics of high energy
collisions and these temperatures occur at different stages. The initial
temperature describes the degree of excitation of an interacting system at
the initial stage of collisions. Chemical freeze out is an intermediate stage
in high energy collisions where the intra-nuclear collisions among the
particles become inelastic and the abundances of stable particles are fixed and the temperature
of the particles at the stage is known as the chemical freeze out temperature.
Correspondingly, the thermal freeze out temperature ($T_0$) describes the excitation
degree of interacting system at the last but not the least stage of high energy
collisions and the final state particles $p_T$ spectra are determined (frozen)
at this stage. Effective temperature is not a real temperature but it describes the sum
of excitation degree of interacting system and the effect of transverse flow at the
stage of kinetic freeze out. Further details of the temperatures can be found in [14].

$T_0$ has very complex situation, based on their energy [15--17] and
centrality dependence [15--19]. Because of having a complex process, the freezeout shows
a hierarchy, where the production of different kinds of particles and reactions cease
at different time scales. According to kinetic theory perspective, the reactions
with lower interaction cross section decouple early from the system compared to
reaction with higher cross section. Besides, the particles decoupling
may also depend on it's rest mass that reveal the multiple kinetic freezeout scenario.
Furthermore, according to some school of thoughts the single, and double kinetic freezeout scenarios also exists in some literatures, which represents one set of parameters to be used for the spectra of both the strange and non-strange particles in case of  single freezeout scenario, while one set of parameters for strange (or multistrange) particles and another for non-strange (or non-multistrange) particles should be used. In case of multiple kinetic freezeout scenario one should use different sets of parameters for different particles. It is very important to find out the correct kinetic freezeout scenario. In addition, $\beta_T$ is also an important
parameter which reflects the collective expansion of the emission source
and it is believed that both $T_0$ and $\beta_T$ maybe dependent on the size of interacting nuclei (A-A,
p-A and p-p collisions) and heavier interacting systems may give the chance of formation of QGP. The
study of various nucleus-nucleus and hadron-nucleus as well as p-p collisions is very useful
in understanding the microscopic features of degrees of equilibration and their dependencies on
the number of participants in the system. Besides, this study may also provide some useful information about the formation of
super hadronic dense matter which is not the focus in this work.

In this paper, our main focal interest is the extraction of kinetic freeze out temperature,
transverse flow velocity and freezeout volume from which we can dig out of the correct kinetic freezeout scenario. We will
extract the relative parameters from the fitting of transverse momentum (mass) spectra of $\pi^+$, $K^+$, $p$, $K^0_S$,
$\Lambda$, $\Xi$ or $\bar\Xi^+$ and $\Omega+\bar \Omega$ or $\bar \Omega^+$ or $\Omega^-+\Omega^+$
produced in central and peripheral Copper-Copper (Cu-Cu), gold-gold (Au-Au) and lead-lead (Pb-Pb)
collisions at 200 GeV, 62.4 GeV and 158 GeV respectively by the blast wave model with
Boltzmann Gibb's statistics. The related parameters are then extracted from the fittings.

The remainder of the paper consists of method and formalism in section 2, followed by the results
and discussion in section 3. In section 4, we summarized our main observations and conclusions.
\\

{\section{The model and method}}

The spectra of high energy transverse
momentum region is generally contributed by the hard scattering process which
is described by the Quantum chromodynamics (QCD) calculations [20--23]. The Hagedorn
function [23,24], which is an inverse power law, can also be used to describe the
hard scattering process. The inverse power law has at least three revisions which
can be found in [25--31].

The low transverse momentum region is contributed by the soft excitation process
in which the kinetic freeze out temperature and transverse flow velocity can be extracted,
however the hard scattering process has no contribution in the extraction of kinetic
freeze out temperature and transverse flow velocity. Therefore we will not discuss
the hagedorn function and it's revisions.

The kinetic freeze out temperature and transverse flow velocity can be extracted by
analyzing the spectra in low $p_T$ region by various distributions such as Erlang
distribution [32--34], Tsallis distribution [35--37], Tsallis+standard distribution[38--43]
and others. We will use the blast wave model with Boltzmann Gibbs statistics [44--46],
which is the most direct distribution and having less parameters. Due to the contribution
of resonance production in some cases or other reasons such as statistical fluctuation.
A single component blast wave model is not enough for the description of spectra in
low transverse momentum region, then the two-component blast wave model is required in
these special cases.

According to references [44--46], the first component of kinetic freeze out temperature
($T_1$) and transverse flow velocity ($\beta_{T1}$) in the blast wave model with
Boltzmann Gibb's (BGBW) statistics results in the probability density function of transverse
momentum ($p_T$) to be
\begin{align}
f(p_T,T_{01}, \beta_{T1})=&\frac{1}{N}\frac{dN}{dp_T} =C_1 \frac{gV}{(2\pi)^2} p_T m_T \int_0^R rdr \nonumber\\
& \times I_0 \bigg[\frac{p_T \sinh(\rho_1)}{T_{01}} \bigg] K_1
\bigg[\frac{m_T \cosh(\rho_1)}{T_{01}} \bigg],
\end{align}
where C$_1$ stands for the normalization constant that leads the integral in Eq. (1) to be normalized to 1, $N$ is the number of particles, $g$ is the degeneracy factor of the particle (which is different for different particles, based on $g_n$=2$S_n$+1, while $S_n$ is the spin of the particle) $m_T=\sqrt{p_T^2+m_0^2}$ is the transverse mass, $I_0$ and $K_1$ are the modified Bessel functions of the first and second kinds respectively, $\rho= \tanh^{-1} [\beta(r)]$ is the boost angle, $\beta(r)= \beta_S(r/R)^{n_0}$ is a self-similar flow profile and $\beta_S$ being the flow velocity on the surface, $r/R$ is the relative radial position in the thermal source.

The second component of the blast wave model has the same form as the first one,
but with the kinetic freeze out temperature ($T_{02}$) and transverse flow velocity
($\beta_{T2}$). thus, the two component transverse momentum distribution function in blast
wave model can be demonstrated as
\begin{align}
f_0(p_T)=kf(p_T,T_{01},\beta_{T1}) +(1-k)f(p_T, T_{02},\beta_{T2}),
\end{align}
where k is the contribution fraction of the first component (soft excitation), while (1-k) shows the contribution fraction of the
second component (hard scattering) in the Eq. 2.
According to Hagedorn thermal model [23], the two-component Boltzmann-Gibbs blast-wave distribution function can also be structured
by using the usual step function,
\begin{align}
f_0(p_T)=\frac{1}{N}\frac{dN}{dp_T}=A_1\theta(p_1-p_T)f(p_T,T_{01},\beta_{T1}) \nonumber\\
+ A_2 \theta(p_T-p_1)
f(p_T,T_{02},\beta_{T2}),
\end{align}
where $A_1$ and $A_2$ are the fraction constants giving the two components to be equal to each other
at $p_T$ = $p_1$, $\theta$ ($p_1$-$p_T$)=1 (or 0), if $p_T$ $<$ $p_1$ (or $p_T$ $>$ $p_1$) and
($p_T$-$p_1$)=1 (or 0), if  $p_T$ $>$ $p_1$ (or  $p_T$ $<$ $p_1$).
Both Eq. (2) and (3) can be used for the extraction of $T_0$ and $\beta_T$ in the two-component
blast wave model i.e
\begin{align}
T_0=kT_{01}+(1-k)T_{02},
\end{align}
and
\begin{align}
\beta_T=k\beta_{T_1}+(1-k)\beta_{T_2},
\end{align}
If Eq.(2) is used to get the parameter values of the two components, k is directly given by
Eq. (2), However, if the parameter of two component values are obtained by using Eq. (3), k is
expressed by
\begin{align}
k=\int_{0}^{p_1} A_1f(p_T,T_{01},\beta_{T1})dp_T,
\end{align}

Because Eq.~(3) is the probability density function, it is naturally normalized.
The second component in Eq. (2) or (3) is not necessary in the fitting procedure. if
the spectra are not in a very wide $p_T$ range, and therefore only the first component
in Eq. (2) and Eq.(3), that is, Eq. (1), can be used to fit the spectra. Although
Eq. (2) and Eq.(3) are not used in this work, but they are presented in order to show
the complete treatment in methodology.

In some cases, the spectra are not in the form of $p_T$, but $m_T$. Then the $p_T$
distribution $f_S$($p_T$) is needed to convert into $m_T$ distribution $f_S$($m_T$) by
$f_S$($m_T$)$|dm_T|$=$f_S$($p_T$)$|dp_T|$ through $p_T$$|dp_T|$=$m_T$$|dm_T|$ due to the invariant
cross section. In fact, Eq. (1) appears in the form of $f_S$($m_T$), so we will convert it into
$f_S$($p_T$) pragmatically. In the present work, we have analyzed the $p_T$ spectra of the
particles ($\pi^+$, $K^+$, $p$, $K^0_S$, $\Lambda$, $\Xi$ or ($\bar\Xi^+$) and $\Omega+\bar \Omega$ or
($\bar \Omega^+$) or ($\Omega^-+\Omega^+$)) by using the single component blast wave model with
Boltzman Gibb's statistics and extracted $T_0$ and $\beta_T$. However the spectra in a wide
$p_T$ range is not needed for the extraction of $T_0$ and $\beta_T$ due to small fraction in high $p_T$ region.
\\

{\section{Results and discussion}}

Figures 1-3 demonstrate the $p_T$ or $m_T-m_0$ or $m_T$ spectra, $(1/N_{ev}) (1/2\pi
m_T)d^2N/dm_Tdy$, $(1/2\pi p_T)d^2N/dp_Tdy$, $(1/m_T)d^2N/dm_Tdy$, $(1/m_T)dN/dm_T$\nonumber\\
of $\pi^+$, $K^+$, $p$, $K^0_S$, $\Lambda$, $\Xi$ or ($\bar\Xi^+$) and
$\Omega+\bar \Omega$ or [$\bar \Omega^+$ or $\Omega^-+\Omega^+$] produced
in central and peripheral Cu-Cu, Au-Au and Pb-Pb collisions at $\sqrt{s_{NN}}$ = 200 GeV,
62.4 GeV and 158 GeV respectively, where $N$ denotes the number of particles.
The types of particle and collision along with their energies are marked in the panels.
The symbols in the figures  represent the experimental data measured by BRAHMS [47],
STAR [48, 45, 49], SPS [50], NA 49 [51] and WA 97 [52] Collaborations respectively. The curves
are our fitted results by using Eq. (1).

The values of the free parameters ($T_0$, $\beta_T$, $V$ and $n$), the normalization constant
($N_0$), $\chi^2$ and the number of degree of freedom (ndof), the concrete collisions, energies,
particles, spectra and the amounts scaled for plotting are given in Table 1. The spectra
in very low-$p_T$ region are not taken care carefully in the fit process due to the resonance
production, while the fit itself is not too good. One can see that the blast-wave fit with Boltzmann
Gibb's statistics fit approximately the experimental data over a wide energy range.

The change in trends of the kinetic freeze out temperature $(T_0)$ and transverse flow
velocity $(\beta_T)$ with the rest mass of the particles $(m_0)$ is demonstrated in figure 4.
Figure 4(a) shows the dependence of $T_0$ on $m_0$ in the central and peripheral Cu-Cu, Au-Au
and Pb-Pb collisions at 200 GeV, 62.4 GeV and 158 GeV respectively for the non-strange,
strange and multi-strange particles, while the dependence of $\beta_T$ on $m_0$ is shown in
figure 4(b). Different collisions are represented by different symbols. The solid and open
symbols represent the central and peripheral collisions respectively.

\begin{figure*}[htbp]
\begin{center}
\includegraphics[width=14.cm]{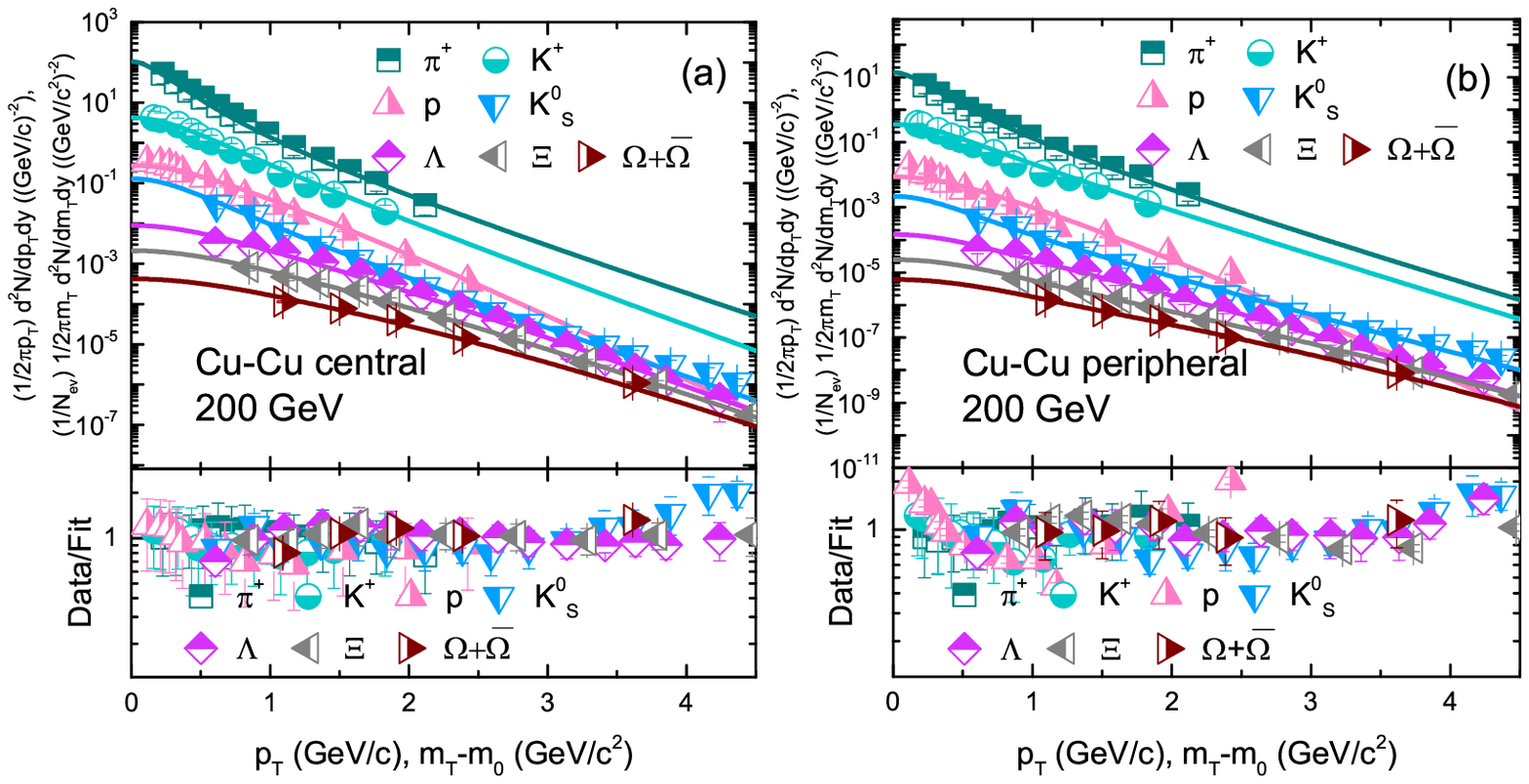}
\end{center}
Fig. 1. Transverse momentum spectra of $\pi^+$,$K^+$ and $p$ at rapidity $|y|=0$ [47], $K^0_S$, $\Lambda$, $\Xi$ and $\Omega+\bar \Omega$ at rapidity $|y|<0.5$ [48] produced in central (0--10\% centrality) and peripheral (50--70\% centrality for $\pi^+$, $K^+$ and $p$ and 40--60\% centrality for $K^0_S$, $\Lambda$, $\Xi$ and $\Omega+\bar \Omega$) Cu-Cu collisions at 200 GeV . The symbols represent the experimental data measured by BRAHMS and STAR collaborations [47, 48], while the curves are our fitted results by using the blast wave model with Boltzmann Gibb's statistics, Eq (1). The corresponding results of data/fit is presented in each panel.
\end{figure*}
\begin{figure*}[htbp]
\begin{center}
\includegraphics[width=14.cm]{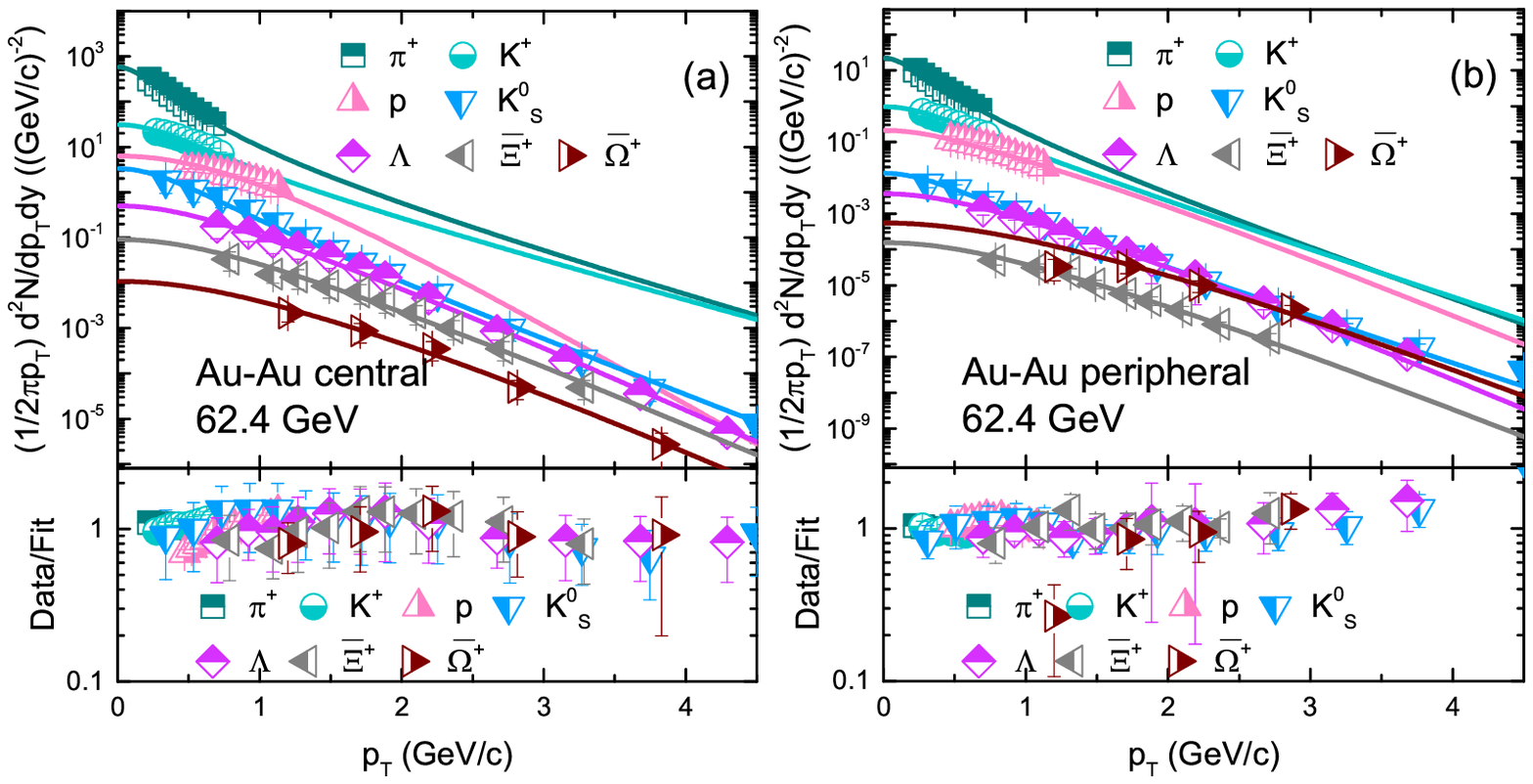}
\end{center}
Fig. 2. Transverse momentum spectra of $\pi^+$, $K^+$ and $p$ at mid-rapidity $mid-|y|<0.1$ [45], $K^0_S$, $\Lambda$, $\bar\Xi^+$ and $\bar \Omega^+$ at pseudo-rapidity $|\eta|<1.8$ [49] produced in central (0--5\% centrality) and peripheral (70--80\% centrality for $\pi^+$, $K^+$ and $p$ and 60--80\% centrality for $K^0_S$, $\Lambda$, $\bar\Xi^+$ and $\bar \Omega^+$) Au-Au collisions at 62.4 GeV. The symbols represent the experimental data measured by  STAR collaboration [45, 49], while the curves are our fitted results by using the blast wave model with Boltzmann Gibb's statistics, Eq (1). The corresponding results of data/fit is presented in each panel.
\end{figure*}
\begin{figure*}[htbp]
\begin{center}
\includegraphics[width=14.cm]{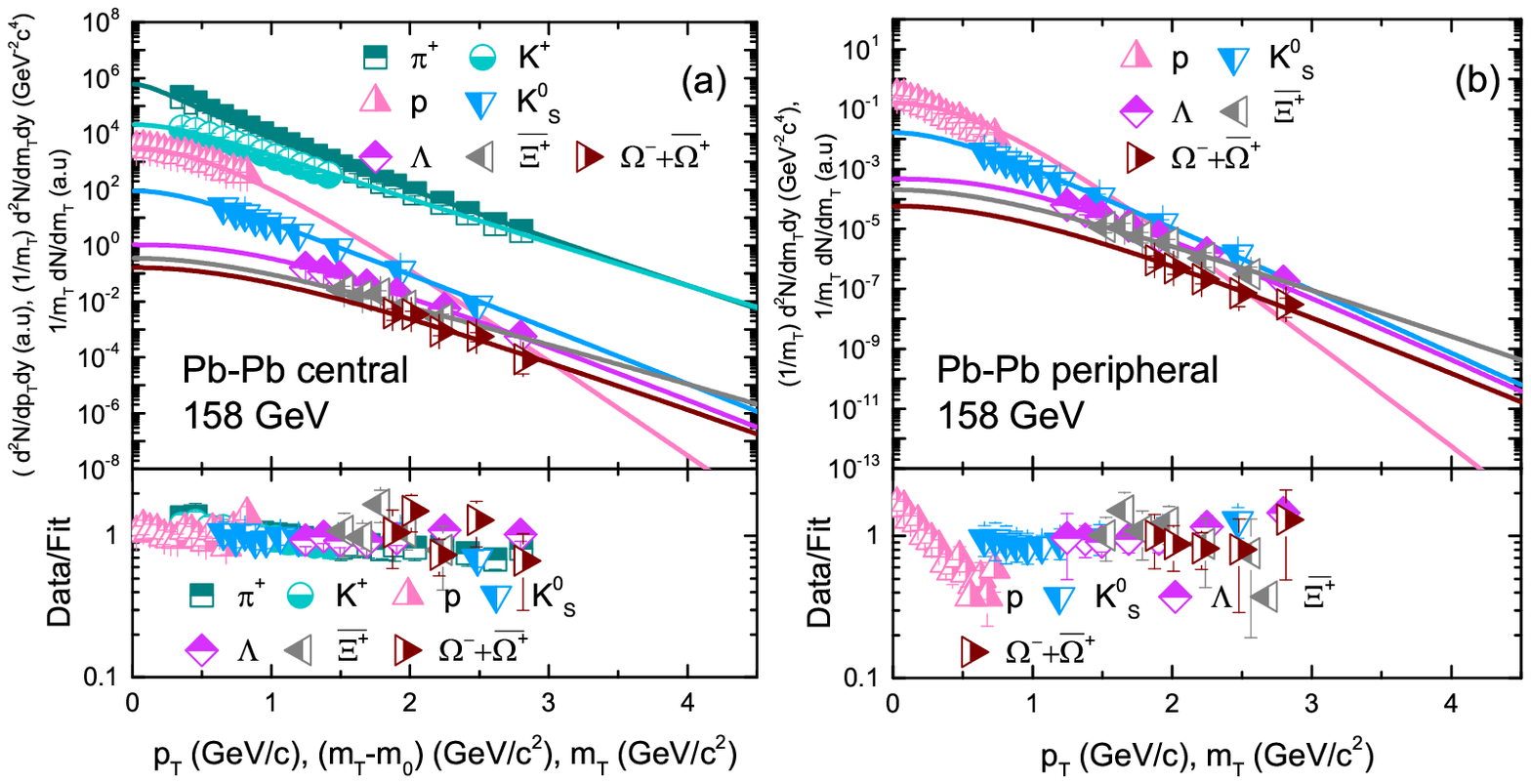}
\end{center}
Fig. 3. Transverse momentum spectra of $\pi^+$,$K^+$ [50], $p$ [51] at rapidity interval $2.4<y<2.8$ and $K^0_S$ and $\Lambda$ at rapidity interval $2 <\eta< 3$, and $\bar\Xi^+$ and $\Omega^-+\bar \Omega^+$ at rapidity interval $3 <\eta< 4$ [52] produced in central and $p$ [51], $K^0_S$, $\Lambda$, $\bar\Xi^+$ and $\Omega^-+\bar \Omega^+$ [52] in peripheral Pb-Pb collisions at 158 GeV . The symbols represent the experimental data measured by SPS, NA 49 and WA 97 collaborations [50--52], while the curves are our fitted results by using the blast wave model with Boltzmann Gibb's statistics, Eq (1). The corresponding results of data/fit is presented in each panel.
\end{figure*}
\begin{figure*}[htbp]
\begin{center}
\includegraphics[width=14.cm]{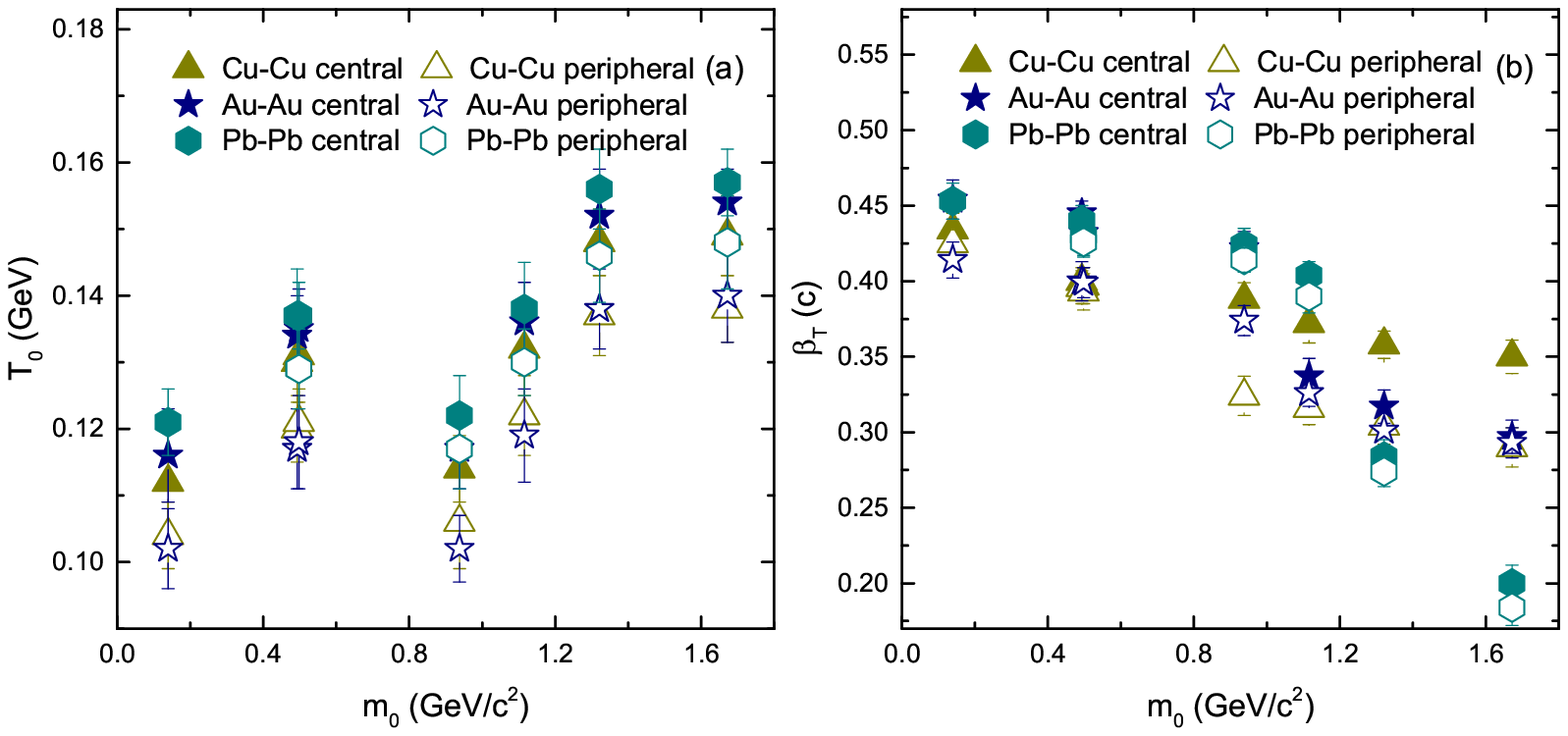}
\end{center}
Fig. 4. Dependence of (a) $T_0$ and (b) $\beta_T$ on $m_0$ and centrality.
\end{figure*}

\begin{table*}
{\scriptsize Table 1. Values of free parameters ($T_0$ and
$\beta_T$), normalization constant ($N_0$), $\chi^2$, and degree
of freedom (dof) corresponding to the curves in Figs. 1--6.
\vspace{-.50cm}
\begin{center}
\begin{tabular}{cccccccccccccc}\\ \hline\hline
Collisions & Centrality & Particle & Spectrum & scaled by & $T_0$ & $\beta_T$ & $V (fm^3)$& $n_0$ & $N_0$ & $\chi^2$ & dof \\ \hline
Fig. 1  & 0--10\% & $\pi^+$ & $1/N_{ev}(1/2\pi m_T)d^2N/dm_Tdy$& -- & $0.112\pm0.008$ & $0.434\pm0.011$& $3490\pm190$ &$1$ & $0.074\pm0.007$ &1 & 9\\
Cu-Cu   & --  &$K^+$&$1/N_{ev}(1/2\pi m_T)d^2N/dm_Tdy$& 100& $0.130\pm0.006$ & $0.400\pm0.009$ &$3000\pm200$&$2$&$1\times10^{-4}\pm3\times10^{-5}$&3& 8\\
200 GeV &  -- & $p$  & $1/N_{ev}(1/2\pi m_T)d^2N/dm_Tdy$ & 2000 &$0.114\pm0.007$ & $0.388\pm0.012$ & $2730\pm155$ &$2$ & $2.7\times10^{-7}\pm4\times10^{-8}$ & 3 & 9\\
        & -- & $K^0_S$ &$(1/2\pi p_T)d^2N/dp_Tdy$& 1/100 &$0.131\pm0.005$ & $0.397\pm0.011$ & $2950\pm163$ &$2$& $0.038\pm0.003$  & 42 & 12\\
        & --& $\Lambda$&$(1/2\pi p_T)d^2N/dp_Tdy$& 1/260 & $0.132\pm0.007$ & $0.372\pm0.013$ & $2500\pm172$ &$2$& $0.0047\pm0.0008$ & 8 & 11\\
        & -- & $\Xi$&$(1/2\pi p_T)d^2N/dp_Tdy$& 1/100 & $0.148\pm0.005$ & $0.358\pm0.009$ & $2170\pm186$ & $2$& $0.0001\pm0.00004$  & 1 & 6\\
        & --& $\Omega+\bar \Omega$& $(1/2\pi p_T)d^2N/dp_Tdy$&  1/100 &$0.149\pm0.006$ & $0.350\pm0.011$ & $1900\pm167$ &$2$ & $1.5\times10^{-4}\pm4\times10^{-5}$&1& 1\\
\cline{2-8}
        & 50--70\% & $\pi^+$  & $1/N_{ev}(1/2\pi m_T)d^2N/dm_Tdy$ & --& $0.102\pm0.005$ & $0.425\pm0.009$ & $2570\pm160$ &$2$& $0.012\pm0.003$ & 2 & 9\\
        &  --  & $K^+$    & $1/N_{ev}(1/2\pi m_T)d^2N/dm_Tdy$& 100 &$0.120\pm0.005$ & $0.396\pm0.011$ & $2300\pm150$ & $2$& $9\times10^{-6}\pm7\times10^{-7}$& 6 & 8\\
        &  --  & $p$      &$1/N_{ev}(1/2\pi m_T)d^2N/dm_Tdy$& 700 &  $0.106\pm0.005$ & $0.324\pm0.013$ & $2000\pm240$ &$2$& $2\times10^{-8}\pm4\times10^{-9}$  & 20 & 9\\
        & 40--60\% & $K^0_S$ & $(1/2\pi p_T)d^2N/dp_Tdy$ &10& $0.121\pm0.007$ & $0.393\pm0.012$ & $2226\pm191$ &$2.3$& $6.5\times10^{-7}\pm5\times10^{-8}$  &32& 12\\
        & -- & $\Lambda$&$(1/2\pi p_T)d^2N/dp_Tdy$&  3& $0.122\pm0.006$ & $0.316\pm0.011$ & $1800\pm140$ & $2.8$& $1.48\times10^{-7}\pm3\times10^{-8}$ & 10 & 11\\
        & -- & $\Xi$&$(1/2\pi p_T)d^2N/dp_Tdy$&7&  $0.137\pm0.006$& $0.304\pm0.012$ & $1600\pm153$ &$3$& $1.7\times10^{-8}\pm4\times10^{-9}$ & 2 & 6\\
        & --& $\Omega+\bar \Omega$&$(1/2\pi p_T)d^2N/dp_Tdy$& 10& $0.138\pm0.005$&$0.290\pm0.013$ & $1400\pm150$ &$3$& $2.12\times10^{-7}\pm2\times10^{-8}$ & 11& 1\\
\hline
Fig. 2  & 0--5\%   & $\pi^+$ &$(1/2\pi p_T)d^2N/dp_Tdy$& -- & $0.116\pm0.007$ & $0.454\pm0.013$ & $5900\pm220$ &$2$ &$0.3\pm0.004$  &11 & 6\\
Au-Au   & --     & $K^+$    &$(1/2\pi p_T)d^2N/dp_Tdy$ & -- & $0.134\pm0.006$& $0.445\pm0.008$ & $5500\pm190$ &$2$ & $0.05\pm0.006$  & 3 & 6\\
62.4 GeV &--    & $p$      & $(1/2\pi p_T)d^2N/dp_Tdy$& 0.5 & $0.117\pm0.006$ & $0.422\pm0.011$ & $5000\pm212$ &$1$ & $0.025\pm0.005$ & 147 & 10\\
        & -- & $K^0_S$ & $(1/2\pi p_T)d^2N/dp_Tdy$& 1/8& $0.135\pm0.006$ & $0.432\pm0.013$ & $5487\pm200$ &$2$ & $0.006\pm0.0005$ & 4 & 10\\
        & -- & $\Lambda$ &$(1/2\pi p_T)d^2N/dp_Tdy$& 1/7& $0.136\pm0.006$ & $0.337\pm0.012$ & $4700\pm200$ & $3$ & $0.0045\pm0.0003$ & 3 & 8\\
        & -- & $\bar \Xi^+$&$(1/2\pi p_T)d^2N/dp_Tdy$& 1/5& $0.152\pm0.007$ & $0.317\pm0.011$ &  $4321\pm213$ &$2$ & $8\times10^{-4}\pm4\times10^{-5}$  & 2 & 6\\
        & -- & $\bar \Omega^+$&$(1/2\pi p_T)d^2N/dp_Tdy$& 1/5&  $0.154\pm0.005$ & $0.297\pm0.011$ & $4000\pm189$ & $2$ & $0.032\pm0.005$  & 1 & 1\\
\cline{2-8}
        & 70--80\% & $\pi^+$  &$(1/2\pi p_T)d^2N/dp_Tdy$& -- & $0.102\pm0.006$ & $0.414\pm0.012$ & $5200\pm198$ &$2$ & $0.008\pm0.0003$ & 3 & 6\\
        &  --    & $K^+$    &$(1/2\pi p_T)d^2N/dp_Tdy$& -- & $0.117\pm0.006$ & $0.400\pm0.013$ & $5000\pm221$  &$2$ & $0.0012\pm0.0004$& 8 & 6\\
        & --    & $p$      &$(1/2\pi p_T)d^2N/dp_Tdy$& 1/2& $0.102\pm0.007$ & $0.374\pm0.010$ & $4700\pm180$ &$2$& $5\times10^{-4}\pm4\times10^{-5}$ & 8 & 10\\
        & 60--80\% & $K^0_S$ & $(1/2\pi p_T)d^2N/dp_Tdy$& $10^4$ &$0.118\pm0.005$ & $0.399\pm0.010$ & $4950\pm168$ &$2$ & $1.6\times10^{-9}\pm3\times10^{-10}$ & 7 & 10\\
        & --& $\Lambda$&$(1/2\pi p_T)d^2N/dp_Tdy$& $10^4$ &  $0.119\pm0.007$ & $0.327\pm0.09$ & $4340\pm165$ &$2$& $5\times10^{-10}\pm6\times10^{-11}$ & 3 & 7\\
        & --& $\bar \Xi^+$&$(1/2\pi p_T)d^2N/dp_Tdy$& $10^3$& $0.138\pm0.006$& $0.301\pm0.012$ & $4000\pm186$ &$2$& $3\times10^{-10}\pm6\times10^{-11}$ & 2 & 5\\
        & --& $\bar \Omega^+$&$(1/2\pi p_T)d^2N/dp_Tdy$& 80&  $0.140\pm0.007$&$0.293\pm0.010$ & $3725\pm152$ &$2$& $9\times10^{-9}\pm4\times10^{-10}$ & 21 & 0\\
\hline \hline
Fig. 3  & Central & $\pi^+$  &$(1/2\pi p_T)d^2N/dp_Tdy$ & -- & $0.121\pm0.005$ & $0.453\pm0.012$ & $7300\pm300$ & $1$ & $169.17\pm32$ & 19 & 18\\
Pb-Pb    & --  & $K^+$  & $(1/2\pi p_T)d^2N/dp_Tdy$ &  -- & $0.137\pm0.007$ & $0.435\pm0.010$ & $7000\pm220$ &$1$ & $23.17\pm6$ & 6 & 8\\
158 GeV & 0--5\% & $p$ & $(1/m_T)d^2N/dm_Tdy$ & 60 & $0.122\pm0.005$ & $0.424\pm0.011$ & $6670\pm190$ &$-2$& $0.02\pm 0.004$ & 15 & 13\\
&  Central& $K^0_S$&$(1/m_T)dN/dm_T$ & 50& $0.137\pm0.006$ & $0.430\pm0.013$ & $6960\pm196$ & $0$ &$0.0017\pm0.0005$ & 10 & 6\\
        & --& $\Lambda$&$(1/m_T)dN/dm_T$& --&$0.138\pm0.007$ & $0.404\pm0.009$ & $6400\pm185$& $0$ & $0.0016\pm0.0007$ & 10 & 3\\
        & --& $\bar \Xi^+$& $(1/m_T)dN/dm_T$& 1/27& $0.156\pm0.006$& $0.284\pm0.011$ & $6113\pm183$ &$2$& $0.014\pm0.004$ & 2 & 1\\
        & -- & $\Omega^-+\bar \Omega^+$&$(1/m_T)dN/dm_T$& 1/180& $0.157\pm0.005$&$0.200\pm0.012$ & $5800\pm200$ &$2$& $0.022\pm 0.003$ & 5 & 1\\
\cline{2-8}
              &43--100\%& $p$ &$(1/m_T)d^2N/dm_Tdy$ & 1/10&  $0.117\pm0.006$ & $0.414\pm0.008$ & $6000\pm210$ & $-2$ & $6\times10^{-5}\pm3\times10^{-6}$ & 117 & 11\\
& Peripheral & $K^0_S$ &$(1/m_T)dN/dm_T$& 20& $0.129\pm0.006$ & $0.426\pm0.010$ & $6280\pm200$ & $0$ & $7\times10^{-7}\pm5\times10^{-8}$ & 3 & 6\\
&  --& $\Lambda$&$(1/m_T)dN/dm_T$& --&$0.130\pm0.005$ & $0.390\pm0.011$ & $5700\pm215$ &$0$ & $7\times10^{-7}\pm3\times10^{-8}$ & 4 & 3\\
        &  -- & $\bar \Xi^+$&$(1/m_T)dN/dm_T$& 1/20& $0.146\pm0.007$& $0.274\pm0.010$ & $5413\pm180$ & $2$& $5\times10^{-6}\pm4\times10^{-7}$ & 2 & 2\\
        & --  & $\Omega^-+\bar \Omega^+$&$(1/m_T)dN/dm_T$&1/300& $0.148\pm0.007$& $0.184\pm0.012$ & $5000\pm210$ &$2$ & $4\times10^{-6}\pm3\times10^{-7}$ & 1 & 1\\
\hline
\end{tabular}%
\end{center}}
\end{table*}

At present, one can see that $T_0$ has no clear dependence on $m_0$. $\pi^+$ and $p$ has lower value for $T_0$ than all of the
strange and multi-strange particles and they decouple from the system at the same time.
$K^+$ and $K^0_S$ are lighter than proton but they both decouple early from the
system and freeze out earlier than proton. This result is inconsistent with [16, 35, 53]
(which shows the obvious mass dependence of $T_0$), but is consistent with [54], although
the main idea in [54] is different from our current work but larger $T_0$ for $K^+$ and $K^0_S$
than proton can be seen.

\begin{figure*}[htbp]
\begin{center}
\includegraphics[width=16.cm]{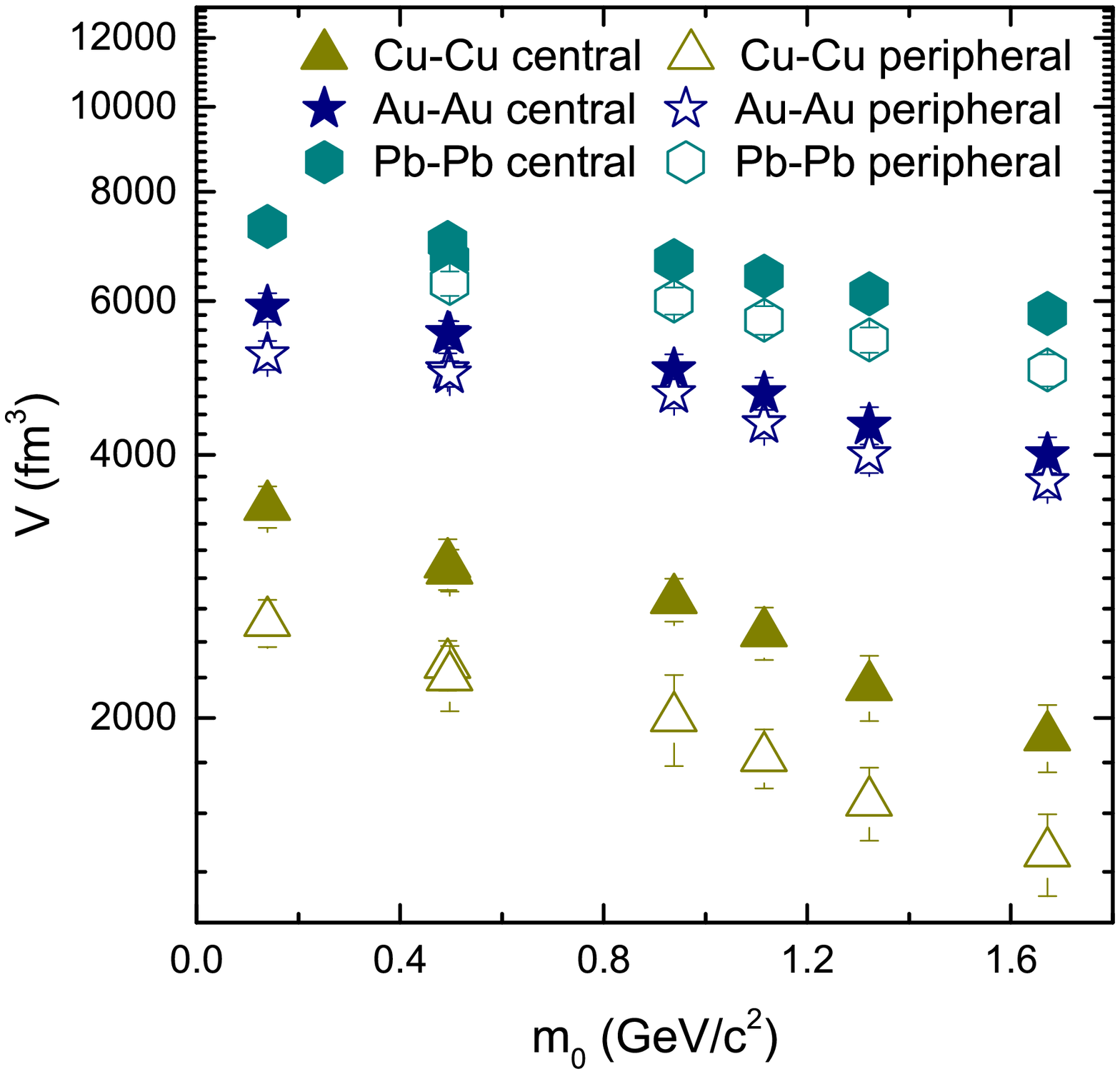}
\end{center}
Fig. 5. Variation of $V$ with $m_0$ and centrality for non-strange, strange and multi-strange particles in Cu-Cu, Au-Au and Pb-Pb  central and peripheral collisions.
\end{figure*}

In the present work, It is observed that the K-F-O temperature of the multi-strange particles
in considerably larger than those of the strange particles and the later is
larger than the non-strange particles, which reveal a picture of separate freeze out processes
for the non-strange, strange and multi-strange particles. We believe that the reason behind the very large kinetic freeze
out temperature of multi-strange particles followed by strange particles  maybe the interaction cross-section,
such that if the multi-strange (strange particles) hadrons don't interact with other hadrons,
their cross-section is small and hence they decouple early from the system, which allow their
early freeze out.

Further more the kinetic freeze out temperature in Cu-Cu, Au-Au and Pb-Pb central collisions
is larger than in peripheral collisions, and $T_0$ for all the particles in Cu-Cu central
and peripheral collisions is smaller than in Au-Au central and peripheral collisions and the later
is smaller than in Pb-Pb central and peripheral collisions, which exhibits the dependence of $T_0$ on
size of interacting system. The larger $T_0$ in the most heaviest nuclei and central collisions is
due to the fact that more number of nucleons are involved in the heavy nuclei and in most central
interactions  which lead the system to higher excitation degree but a variety of data maybe needed
(hadron-hadron, hadron-nucleus, different nucleus-nucleus collisions) to further analyze
this work in the future.

Figure 4(b) shows the variation of of $\beta_T$ with $m_0$. One can see that $\beta_T$ is slightly larger
in central collisions than in the peripheral collisions. Furthermore, it is observed that $\beta_T$  decrease with
increasing mass of the particle which indicates the early decoupling of heavier particles from the
system that results in their early freeze out of heavy particles. This result is consistent with [16, 18, 55].
No dependence of $\beta_T$ on energy or size of the interacting system is observed in this work.

Figure 5 is the same as Fig4. but it demonstraes the variation of $V$ with $m_0$ and centrality for the non-strange, strange and multi-strange particles in Cu-Cu, Au-Au and Pb-Pb central and peripheral collisions at 200, 62.4 and 158 GeV respectively. Different symbols are used to represent different particles, while the solid and open symbols represent the central and peripheral collisions respectively. One can see the larger $V$ for light particles which decrease for heavier particles that leads to the volume differential freezeout scenario. Compared to the peripheral collisions, the central collisions correspond to larger larger $V$ due large number of participant nucleons which indicates the quick approach of the later system to equilibrium state. Furthermore, $V$ is observed to be dependant on the size of the interacting system as it is larger in Pb-Pb collisions than in Au-Au collisions, and in later it is larger than in Cu-Cu collisions.

\begin{figure*}[htbp]
\begin{center}
\includegraphics[width=16.cm]{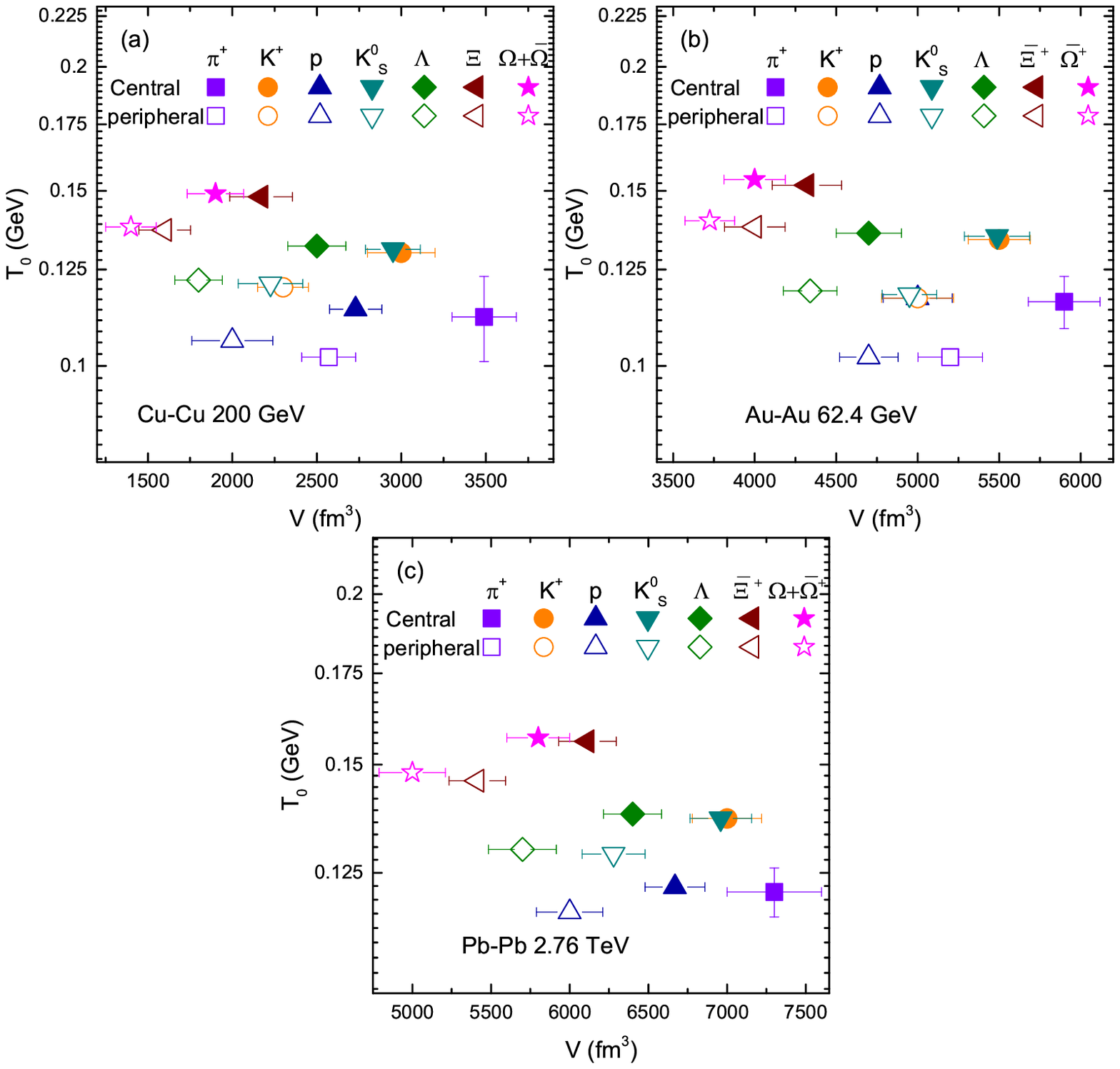}
\end{center}
Fig. 6. Variation of $T_0$ with $V$ for non-strange, strange and multi-strange particles in Cu-Cu, Au-Au and Pb-Pb  central and peripheral collisions.
\end{figure*}

\begin{figure*}[htbp]
\begin{center}
\includegraphics[width=16.cm]{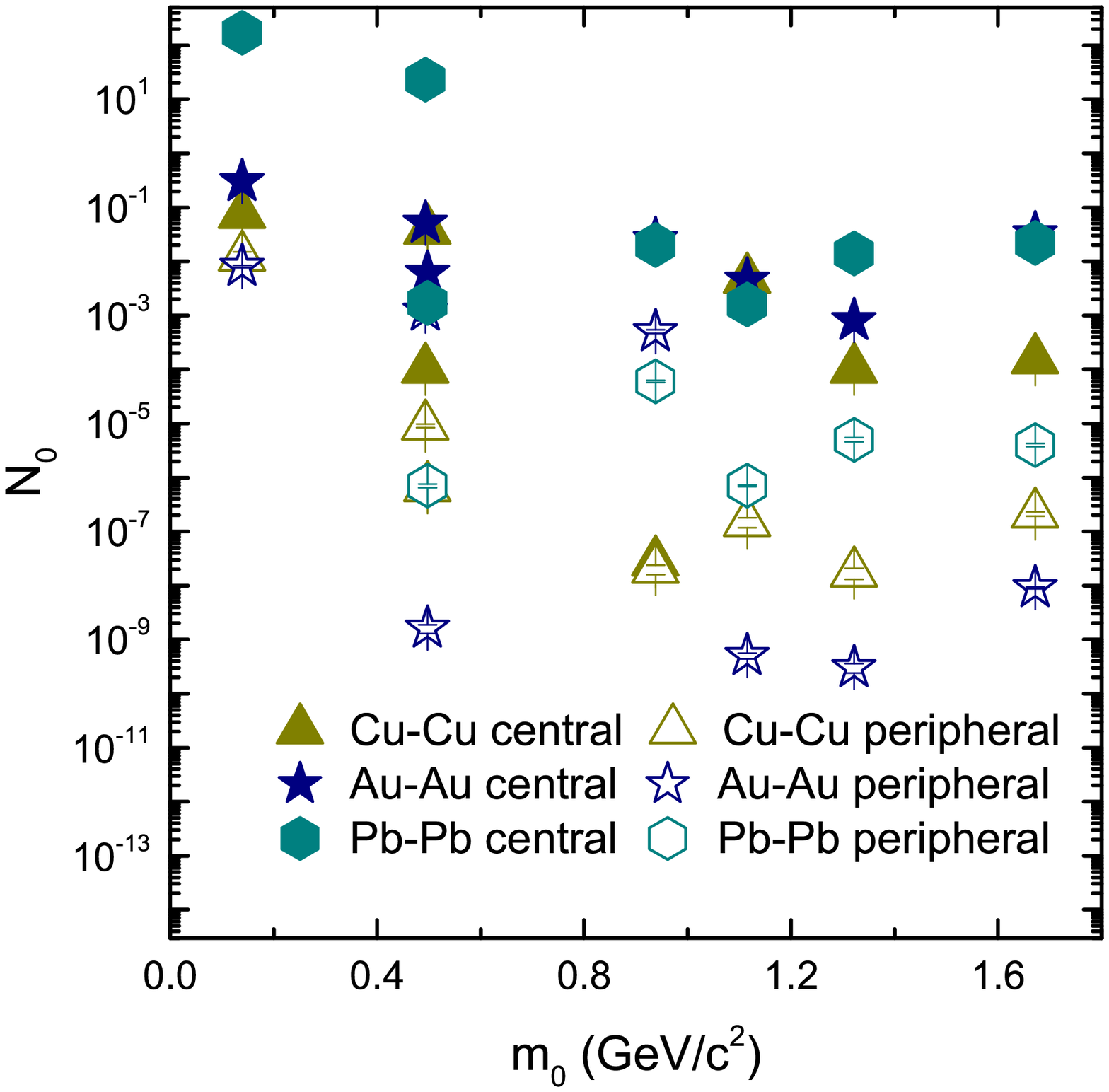}
\end{center}
Fig. 7. Variation of $N_0$ with $m_0$ and centrality for non-strange, strange and multi-strange particles in Cu-Cu, Au-Au and Pb-Pb  central and peripheral collisions.
\end{figure*}

Figure 6 (a)-(c) shows the co-relation of $T_0$ and $V$. The solid and open symbols are used for the central and peripheral collisions respectively and different particles are represented by different symbols. $T_0$ decrease with the increase of $V$ in the most central and most peripheral heavy ion (Cu-Cu, Au-Au and Pb-Pb) collisions.

Figure 7 is the same to Fig 4. but it demonstrates the dependence of $N_0$ with $m_0$ and centrality. $N_0$ is not just a normalization constant but it reflects the multiplicity. One can see lager $N_0$ in central collisions compared to the peripheral collisions. At present, there is no dependence of $N_0$ on $m_0$.
\\

{\section{Conclusions}}

We summarize here our main observations and conclusions.

(a) The transverse momentum spectra of non-strange, strange and multi-strange particles
produced in central and peripheral Cu-Cu, Au-Au and Pb-Pb collisions have been analyzed by BGBW model.
The model results are in agreement with the experimental data in the special
$p_T$ range measured by BRAHMS, STAR, SPS, NA 49 and WA 97 collaborations.

(b) Separate kinetic freezeout scenario for non-strange, strange and multi-strange particles are found which shows the dependence
of the kinetic freezeout temperature of the particles on their interaction cross-section and reveals the triple kinetic freeze out scenario, However $\beta_T$ and $V$ are mass dependent which decrease with increasing $m_0$.

(c) Kinetic freeze out temperature ($T_0$), transverse flow velocity ($\beta_T$) and kinetic freezeout volume ($V$) are extracted
from the transverse momentum spectra fitting to the experimental data. It is observed that $T_0$, $\beta_T$ and $V$ are slightly larger in central collisions than in the peripheral collisions, and $T_0$ and $V$ is also generally larger in the most heaviest interacting system, while $\beta_T$ shows no dependence on the size of interacting system.

d) The normalization constant ($N_0$) that reflects multiplicity, is larger in central collisions than in peripheral collisions.
\\
{\bf Data availability}

The data used to support the findings of this study are included
within the article and are cited at relevant places within the
text as references.
\\
\\
{\bf Compliance with Ethical Standards}

The authors declare that they are in compliance with ethical
standards regarding the content of this paper.
\\
\\
{\bf Acknowledgements}

The authors would like to thank support from the National Natural Science
Foundation of China (Grant Nos. 11875052, 11575190, and 11135011).

\end{multicols}
\end{document}